\documentstyle[11pt,newpasp,twoside,epsf]{article}

\def\spose#1{\hbox to 0pt{#1\hss}}
     
\begin{document}

\title{An Overview of the Las Campanas Distant Cluster Survey}

\author{Dennis Zaritsky}
\affil{Steward Observatory, 933 N. Cherry Ave., University of Arizona,
Tucson, AZ, 85721, USA}

\author{Anthony H. Gonzalez}
\affil{Harvard-Smithsonian Center for Astrophysics, 60 Garden St., Cambridge, MA, 02138, USA}

\author{Amy E. Nelson}
\affil{Department of Astronomy and Astrophysics, Univ. of Calif. at Santa Cruz,
Santa Cruz, CA, 95064, USA}

\author{Julianne J. Dalcanton}
\affil{Department of Astronomy, University of Washington, Box 351580, Seattle, WA, 98195, USA}

\begin{abstract}
We present the Las Campanas Distant Cluster Survey, which has produced
over a thousand galaxy cluster candidates at $0.35 < z < 1.1$ (see Gonzalez
{\it et al.} 2001 for the full catalog). 
We discuss the technique that enabled us to use short 
($\sim 3$ min) exposures
and a small (1m) telescope to efficiently survey $\sim$ 130 sq. deg.
of sky. Follow-up imaging and spectroscopy using a wide array of telescopes
including the Keck and VLT suggest that the bona-fide cluster fraction
is $\sim$ 70\%. We construct methods to estimate both the redshift
and cluster mass from the survey data themselves and discuss our
first result on large-scale structure, the dependence of the cluster
correlation length with mean cluster separation at $z \sim 0.5 $
(Gonzalez, Zaritsky, \& Wechsler 2001).
\end{abstract}


\section{Introduction}

Each of the many methods with which to find the most massive, 
gravitationally-relaxed objects
in the universe has its own relative advantages and disadvantages. 
As discussed by Yee (this volume), the surveys
by Abell and Zwicky pioneered this field decades ago.
Current surveys exploit
differing wavelengths (from submm to X-ray) and techniques (from identifying
an excess of galaxies to identifying an excess of mass). Because galaxy
clusters are not idealized, isolated, fully-relaxed systems, complementary
techniques are necessary to ensure
that all potential systematic difficulties introduced by the particular
survey method are identified and explored.
We have introduced
a method that enables us to identify high redshift galaxy cluster 
candidates using modest exposures on small telescopes (Dalcanton 1996;
Zaritsky {\it et al.} 1997; Gonzalez {\it et al.} 2001).

\section{The Survey}

The basic premise of our detection technique is that we utilize
the light from unresolved cluster galaxies. Rather than 
obtaining deep images of the sky in order to detect a statistically
significant number of cluster {\it galaxies}, 
we only need to obtain an image that
contains a statistically significant number of cluster {\it photons}. 

Our current survey is based on drift scan observations obtained with
the Las Campanas 1m Swope telescope and the Great Circle Camera
(Zaritsky, Shectman, \& Bredthauer 1996) over a $\sim$ 130 sq. deg.
area of sky. We obtained two scans through
each region of the survey area, each with an effective exposure time
of $\sim$ 90 s. The power of the technique is manifested by the detection
of clusters out to $z \sim 1.1$ with such shallow data (see below).
Briefly, the reduction and analysis involves several flat-fielding passes
(to remove CCD response variations and sky fluctuations), 
masking of bright stars, removal of resolved galaxies and faint stars, 
and smoothing with a kernel that corresponds roughly to the size of
cluster cores at $z \sim 0.6$. Statistically significant low surface
brightness (LSB) fluctuations of the correct character are cluster candidates.
We refer the interested reader to Gonzalez {\it et al.} (2001) for
a full description of the data reduction techniques and the candidate
cluster catalog.

\begin{figure}
\plotfiddle{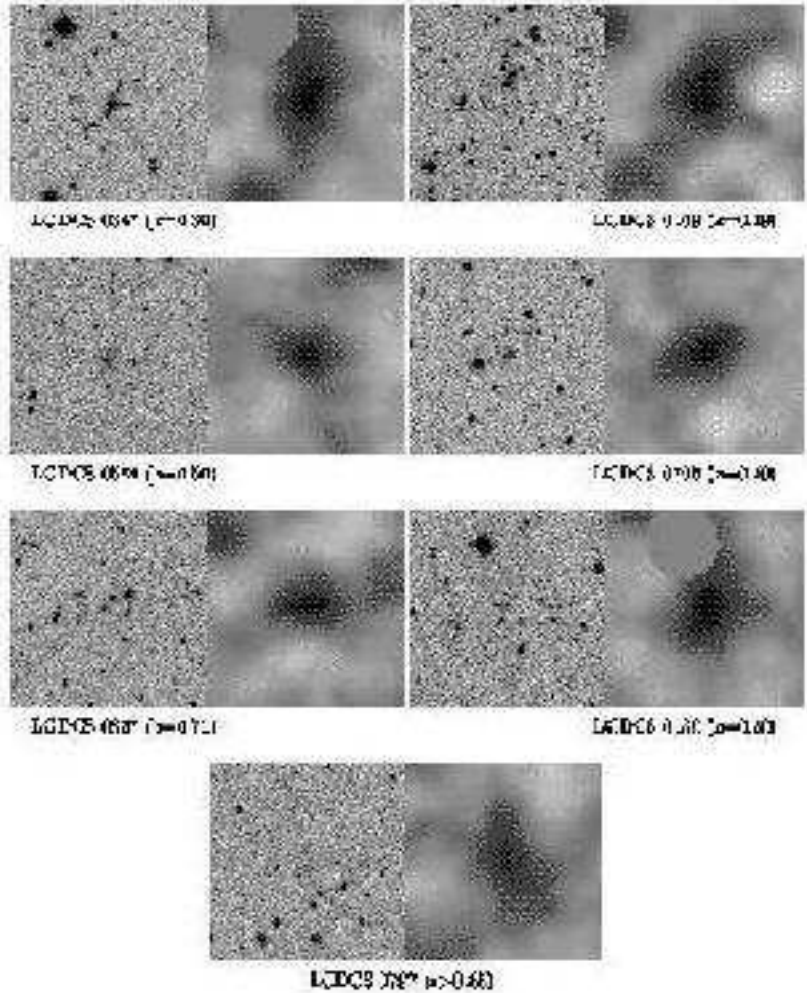}{320 truept}{0}{60}{60}{-170}{-20}
\caption{Examples of candidate clusters. Left panel shows the
original survey data, right panel has the cleaned and smoothed
version. Estimated redshifts range from 0.3 for the upper left to
$>$ 0.85 for the lowest panel. The intermediate step where resolved
galaxies and stars are removed is not shown. The LSB fluctuation
does not arise from the resolved sources visible in the left panels.}
\end{figure}

\section{Follow-Up Observations}

The telescope-intensive part of the project has been the follow-up
observations necessary to 
test our cluster candidates and calibrate the 
methods we are using to estimate the cluster redshift and mass (see below). 
Such expensive follow-up methods are common to all surveys regardless
of whether they originate from optical, X-ray, or SZ data. We now describe some of 
the observations that lead us to conclude that the overall
contamination of the cluster catalog is $\sim$ 30\% (with greater
contamination toward higher redshifts). While X-ray and SZ surveys
should have lower contamination rates, optical follow-up will still be 
necessary at least for redshift determination.

\subsection{Photometry}

A basic method to confirm cluster candidates is to obtain deeper images
that enable one to identify a concentration of galaxies at the 
position of the LSB fluctuation
and a red galaxy sequence in color-magnitude diagrams
characteristic of early-type galaxies in clusters. We show examples
of these two approaches using data presented by Nelson {\it et al} (2001).
First, in Figure 1 we plot the radial density of galaxies in 
cluster candidates that we have deemed to be bona-fide. There is a 
clear central concentration of galaxies indicative of a cluster.
Second, in Figure 2 we plot the color of the red sequence vs. 
spectroscopic redshift (see below) for clusters from our survey vs.
clusters from the literature (some of our cluster candidates come from
an exploratory survey done using Palomar 5m drift scans; Dalcanton {\it 
et al.} 1997). The excellent agreement between the 
two samples is indicative
that the objects that we call clusters (not all candidates, but rather
those $\sim$ 70\% that we deem to be bona-fide clusters) are indeed
similar to cluster in the literature. Finally, we plot
the results from recent VLT observations for an ongoing program
that aims to investigate the detailed properties of 10
clusters at $z \sim 0.5$ and another 10 at $z \sim 0.8$. These
data come from the initial snapshots intended to confirm clusters
candidates before more observing time is spent. Here we include 
all of the $z \sim 0.5$ candidates (regardless of whether in the
final analysis we deem them to be bona-fide). Red sequences are 
prominent in the majority, again confirming that the contamination 
rate is not significantly larger than 30\%.

\begin{figure}
\plotfiddle{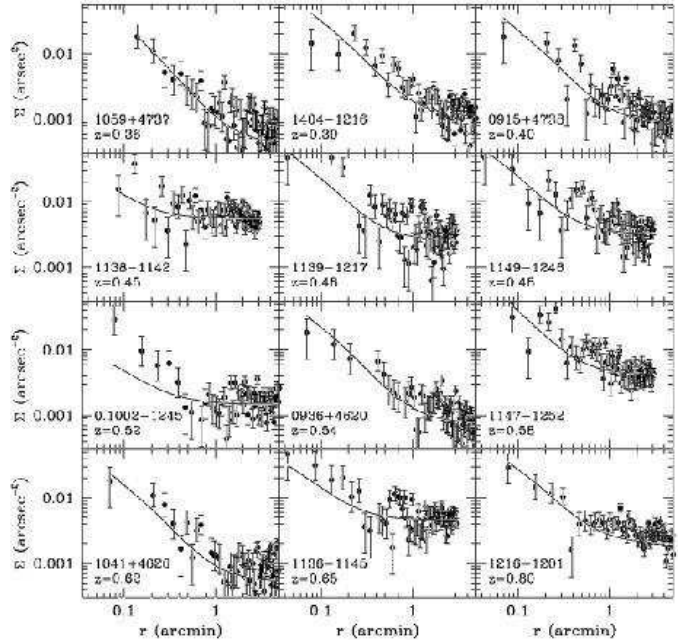}{230 pt}{0}{50}{50}{-140}{-30}
\caption{Radial profiles of a sample of ``confirmed" cluster candidates.}
\end{figure}

\subsection{Spectroscopy}

Complementary confirming evidence comes from spectroscopy. 
In Figure 4 we present all of our spectroscopic follow-up data
from the Keck telescopes. The spectrograph slit was placed
at the position of the LSB feature identified
in the original survey data and the guider was used to rotate
the slit in such a way as to include as many galaxies as possible.
The groupings in redshift space 
of three or more galaxies that resulted from these observations
in the majority of the cluster candidate fields
again confirms our contamination rate. Monte-Carlo simulations using
Keck {\it field} redshift surveys to similar magnitudes suggest that
our spectroscopic sample could include one random three galaxy grouping and
no random four galaxy groupings. Some of the failed cluster candidates
may actually be clusters because 1) we may have been unfortunate in our
placement of the spectrograph slit and simply missed including
enough cluster galaxies, and 2) some of the failed candidate fields
received substandard exposures (due to time constrains or weather).

\begin{figure}
\plottwo{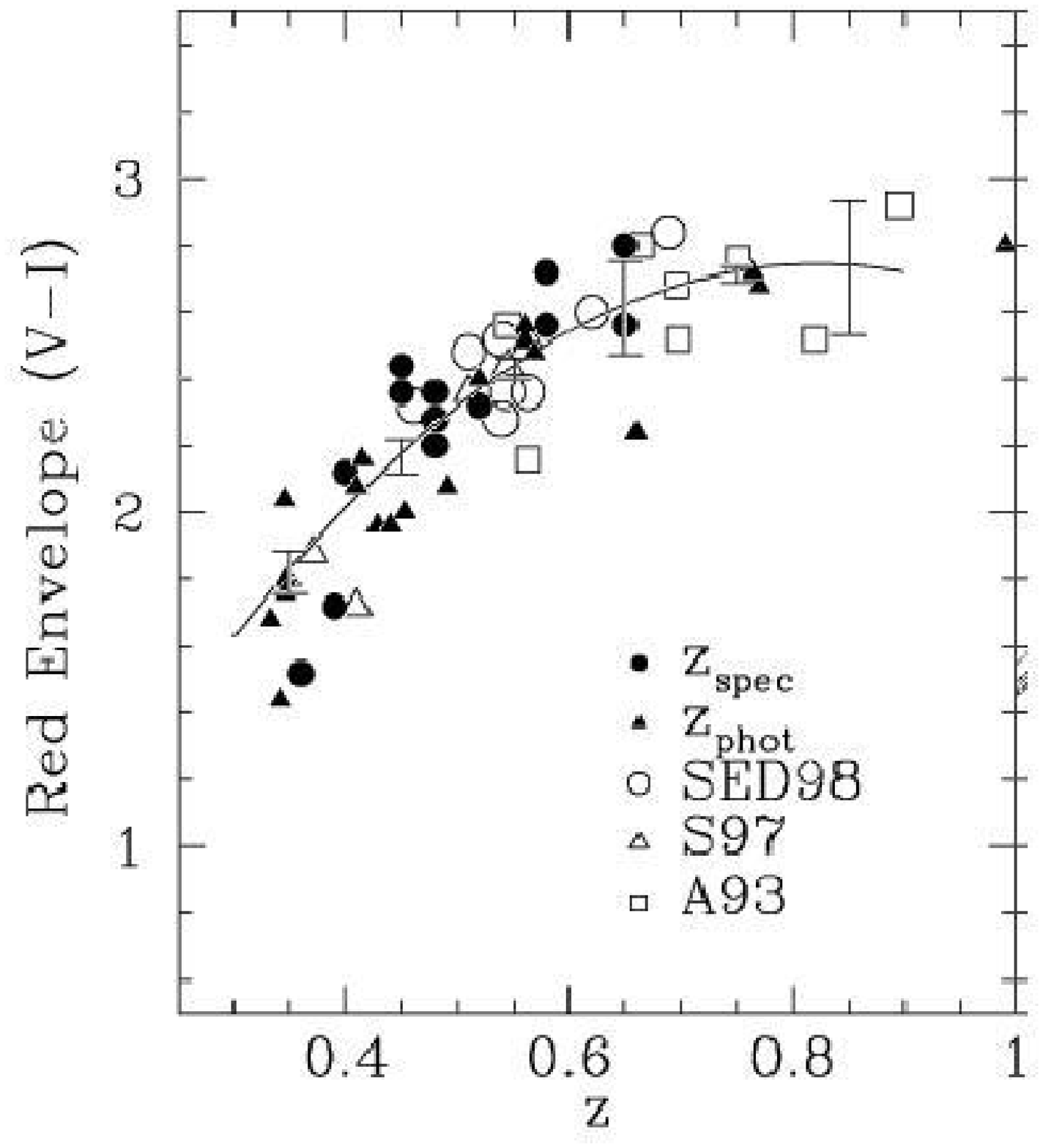}{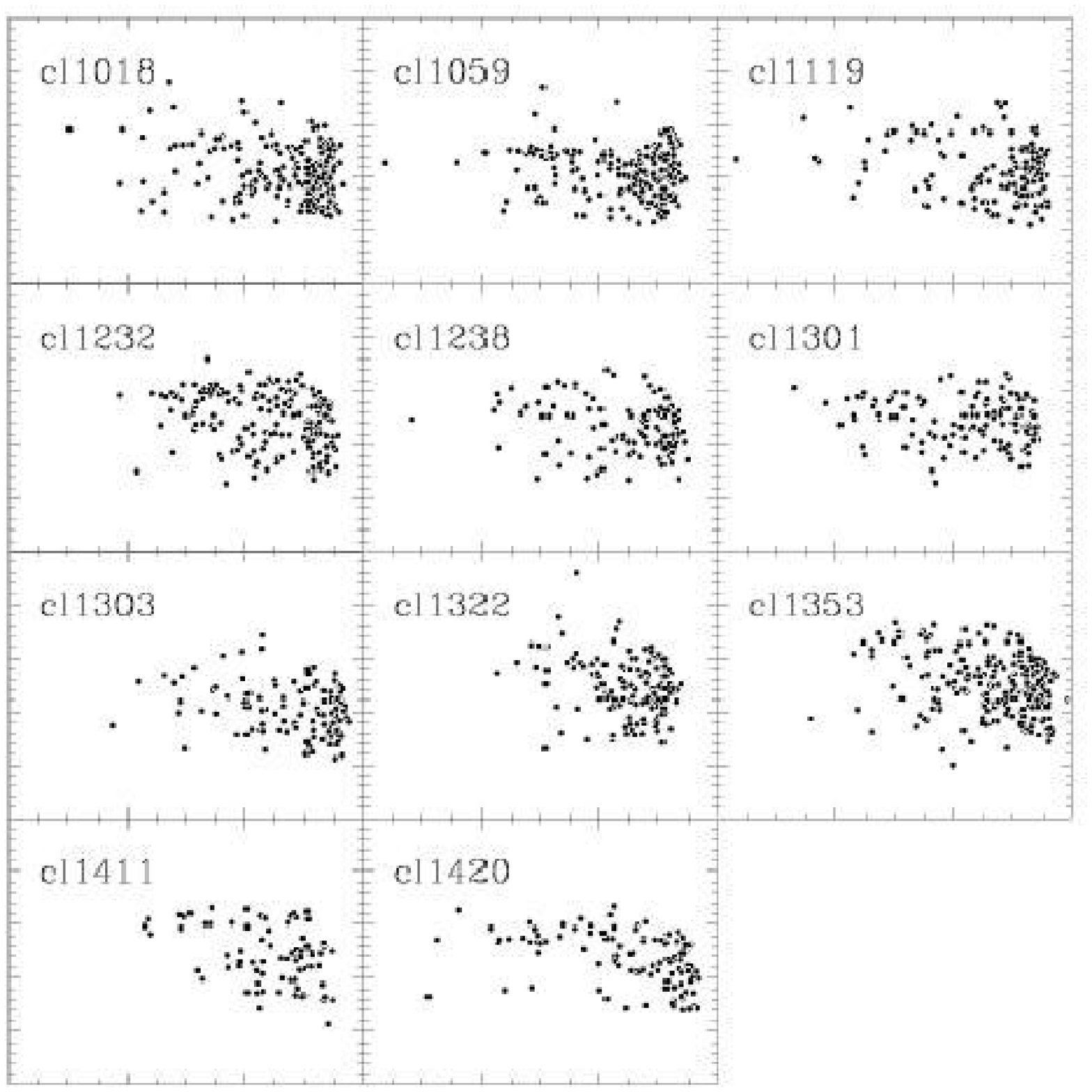}
\caption{Photometric confirmation. Left panel plots the color of the
red sequence for our clusters (solid symbols) vs. literature cluster (open
symbols).
The line represents an  empirical fit the the color vs. redshift
relation (see Nelson {\it et al.} 2001 for a full description). The right panel shows color-magnitude diagrams from VLT snapshot
exposures of $z \sim 0.5$ candidate clusters.}
\end{figure}

\begin{figure}
\plotone{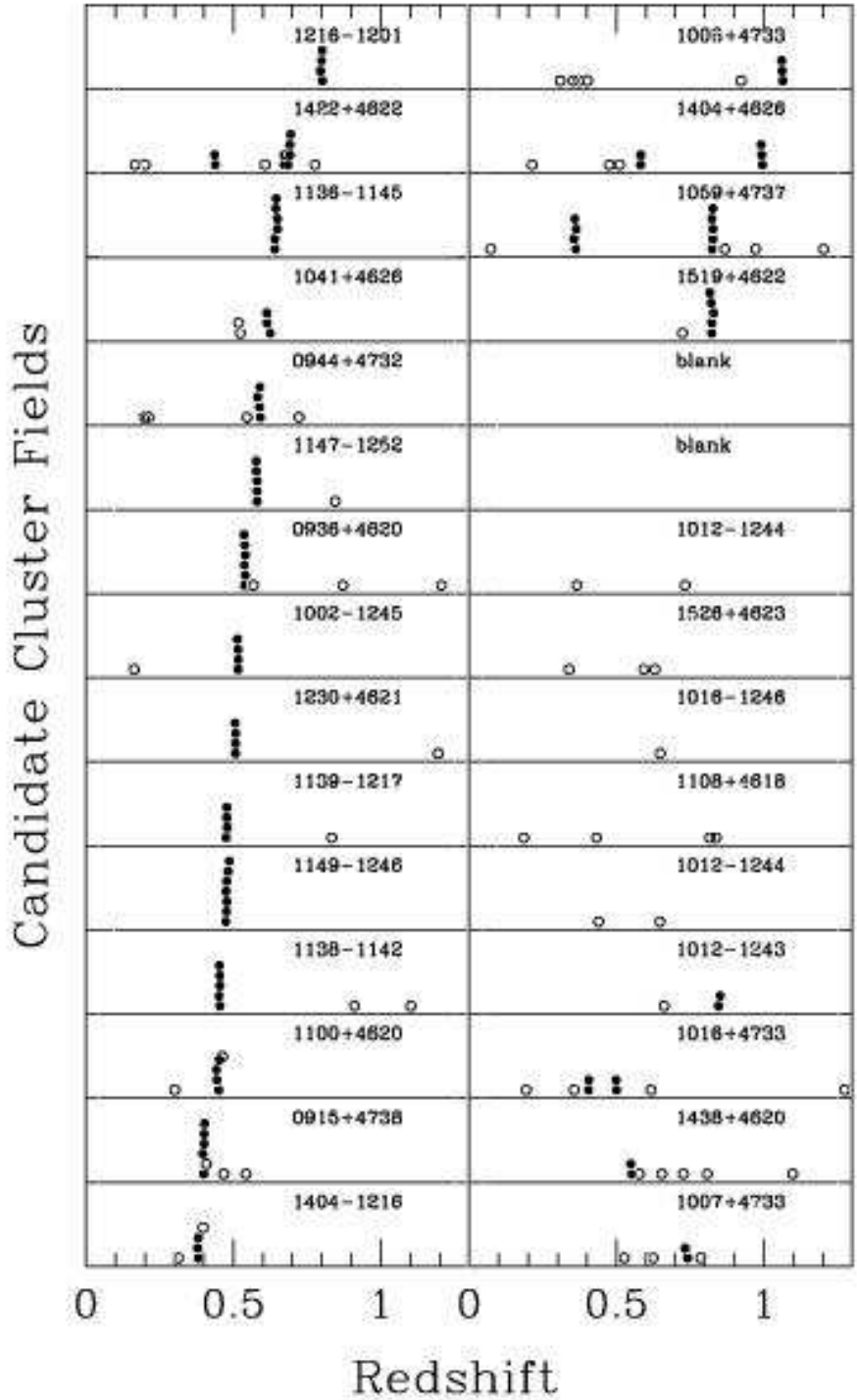}
\caption{Spectroscopic follow-up of cluster candidates. Each panel
represents one cluster candidate field. In each panel, the filled
circles represent galaxies within 1000 km sec$^{-1}$ of another galaxy
and open circles represent other galaxies. The two empty panels
are not candidate cluster fields. The candidates are sorted into
two groups (likely clusters and unlikely clusters) and within the
former group by redshift.}
\end{figure}

\section{Calibrating Redshift and Mass Diagnostics}

In surveys that produce thousands (or even hundreds) of galaxy
cluster identifications, it is impractical to obtain redshift
and mass measures via spectroscopy for a significant fraction
of the candidates. Ideally, these quantities should
be estimated from the survey data themselves. As such, we clearly
sacrifice precision on a cluster-by-cluster basis, but if
the uncertainties are well understood, the sheer number of clusters 
allows high precision measures of the statistical properties of the sample. 
There really is no choice in this matter (in this survey or future
SZ surveys). For example, to obtain a sufficient number of 
spectroscopic redshifts of cluster galaxies
for a reliable velocity dispersion measure ($\sim$ 50) 
in a sample of ONLY 20 clusters we require $\sim$ 35 VLT nights!

\begin{figure}
\plotfiddle{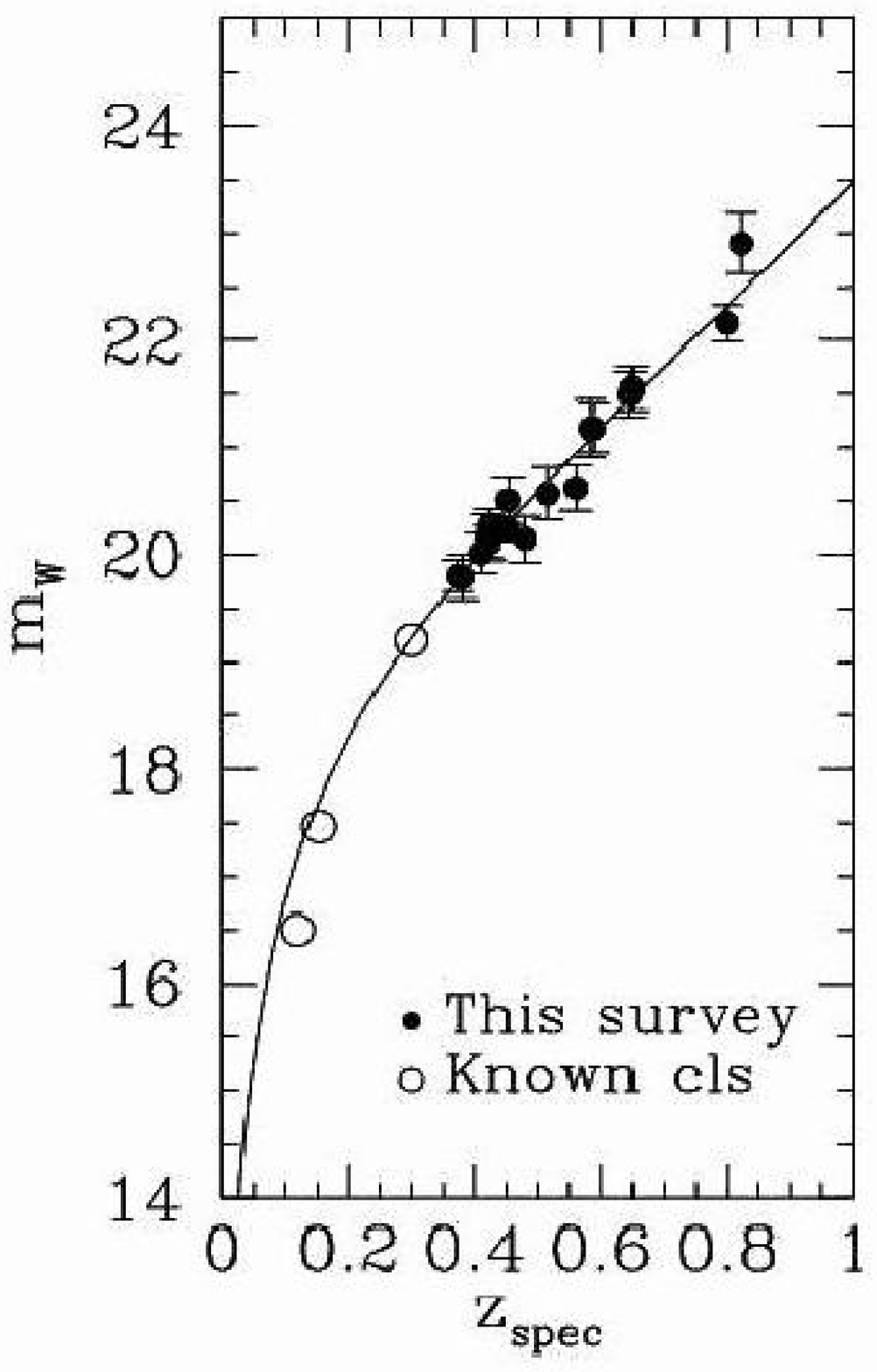}{200 pt}{0}{35}{28}{-110}{-15}
\caption{BCG magnitudes vs. redshifts. The open circles represent
known foreground clusters that lie within our survey. $m_W$ is the 
galaxy magnitude for the wide optical filter used in the survey.}
\end{figure}

We use the magnitude of the brightest cluster
galaxy (BCG) as a redshift indicator and the central surface brightness
of the smoothed cluster detection ($\Sigma$) as a mass indicator. BCGs are 
excellent standard candles locally (Humason, Mayall, \& Sandage 1956;
Graham {\it et al.} 1996)
and at high redshift (Aragon-Salamanca {\it et al.} 1993.). 
In Figure 5 we show 
our empirical calibration of the BCG magnitudes in our photometric
system vs. spectroscopic redshift (as obtained for the cluster from
the Keck data). The dispersion is $\sim 0.08$ in $z$ once a
correction is applied to the BCG magnitude for cluster mass (again
empirically calibrated from the data). However, the redshift error distribution
is non-Gaussian and we construct the full distribution by randomly 
inserting cluster candidates into the original survey images 
and reanalyzing them (see
Gonzalez {\it et al.} (2001) for a full description).

The calibration of the mass indicator is much more speculative
primarily because of the dearth of independent mass estimates for high redshift
clusters. Only one previously known X-ray cluster at $z > 0.35$ lies
within our survey area, so we obtained small drift scans around 17 other
clusters with published X-ray luminosity and/or temperature measurements. 
Nevertheless, the relationships between $\Sigma$ and other mass
indicators ($L_x, T_x,$ or velocity dispersion) are poorly defined.
Two aspects are particularly vexing: 1) we need to determine not
only the relation between these quantities but we also need to 
quantify the scatter, and 2) most
of the data are for lower redshift clusters, complicating 
the removal of redshift-dependent effects like evolution.
Extracting the full potential of this, or any other
survey, is predicated on developing a reliable mass estimator and
understanding its uncertainties.

\section{The Correlation Function of Clusters}

As an example of statistical results that can be obtained from
our cluster catalog we briefly discuss our first result 
regarding large-scale structure. The cluster-cluster correlation
function has a complicated history that cannot be outlined here.
However, it is generally parametrized by examining the behavior of
the correlation scale-length, $r_0$, vs. the mean cluster
separation of a sample, $d_c$. The latter is a measure of the
mass of the clusters because more massive clusters are rarer and
so have larger mean cluster separations. 

\begin{figure}
\plottwo{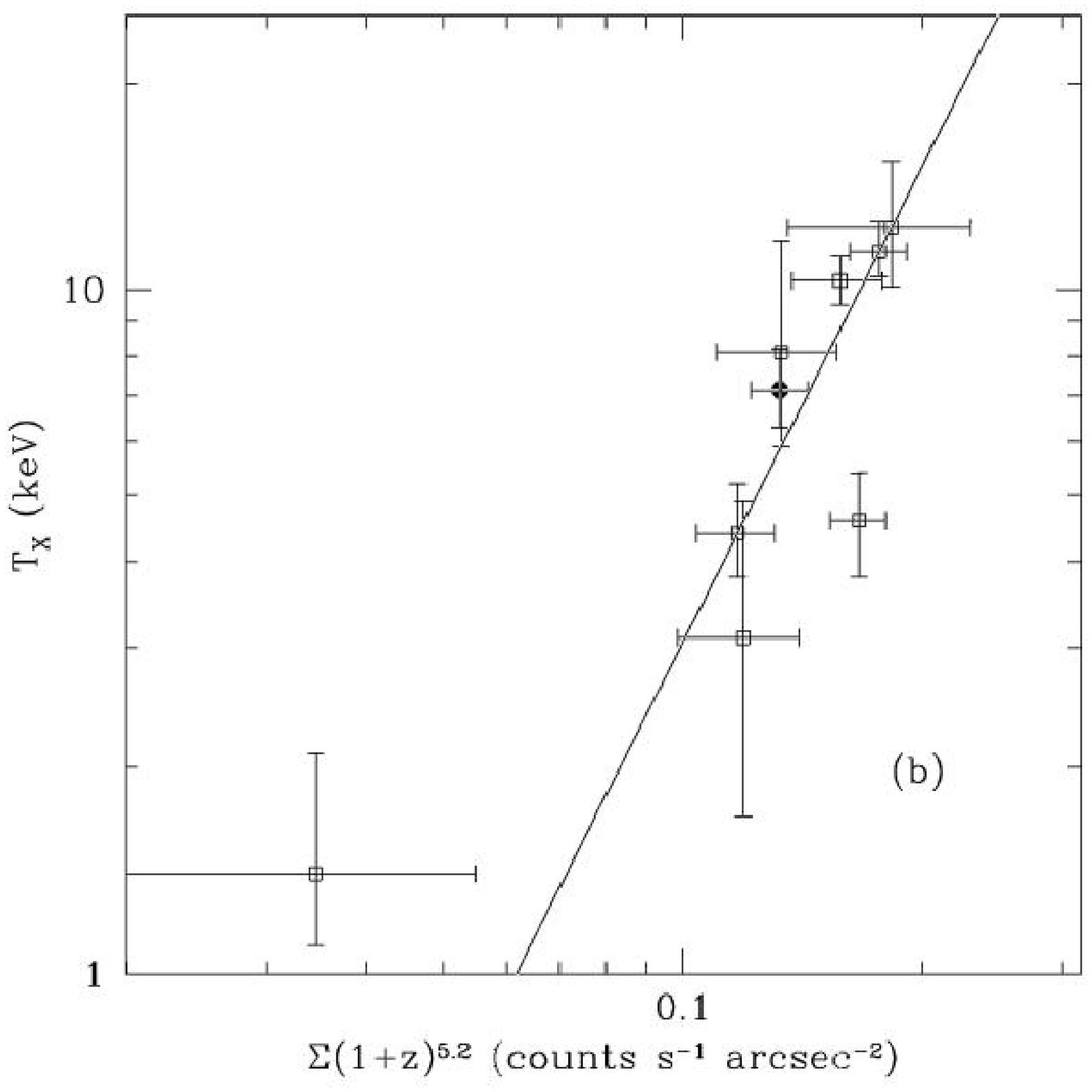}{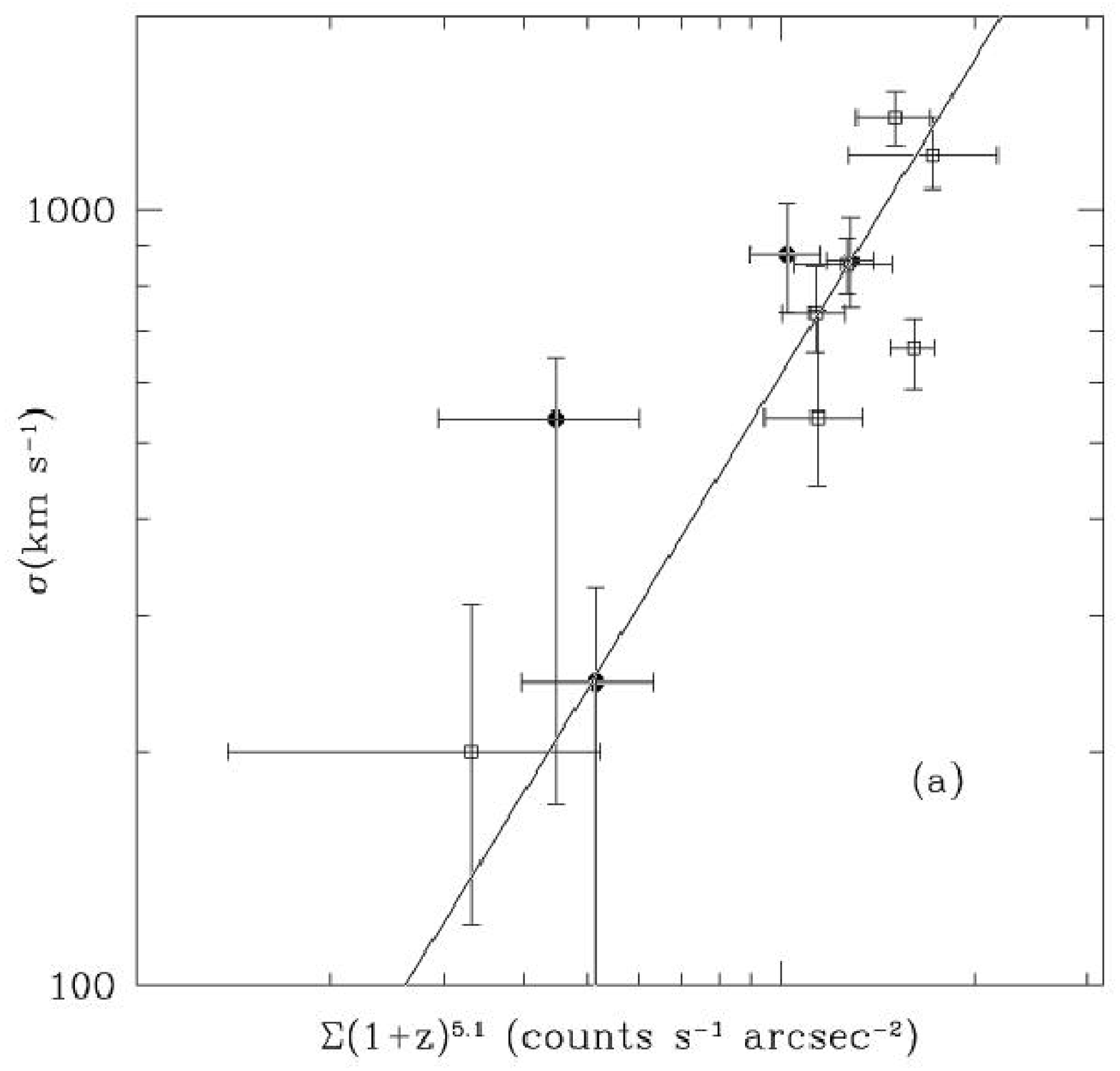}
\caption{X-ray temperature and galaxy velocity dispersion vs. 
$\Sigma$ (corrected for cosmological dimming and evolution empirically)}
\end{figure}

We select a subsample in the redshift range where we believe
our contamination is lowest and best understood ($ 0.35 < z < 0.6$),
and for which our redshift and mass estimates are more robust.
We invert the Limber equation to convert the angular correlation
function into a spatial function. The comparison of our results
with those from local ($z < 0.2$) surveys and simulations is
presented in Figure 7. We find that the
correlation of clusters at $z \sim 0.5$ is quite similar to that
seen at lower redshifts. The low redshift results agree with the
predictions from the VIRGO consortium simulation (Colberg {\it et al.}
2000) and with the very moderate
amount of evolution expected theoretically. This measure
of large-scale structure is not a particularly powerful 
discriminator amongst the currently allowed cosmological models, 
but the results demonstrate that our measured clusters at intermediate
redshifts are consistent with our current theoretical understanding.

\section{Conclusions}

\begin{figure}
\plotfiddle{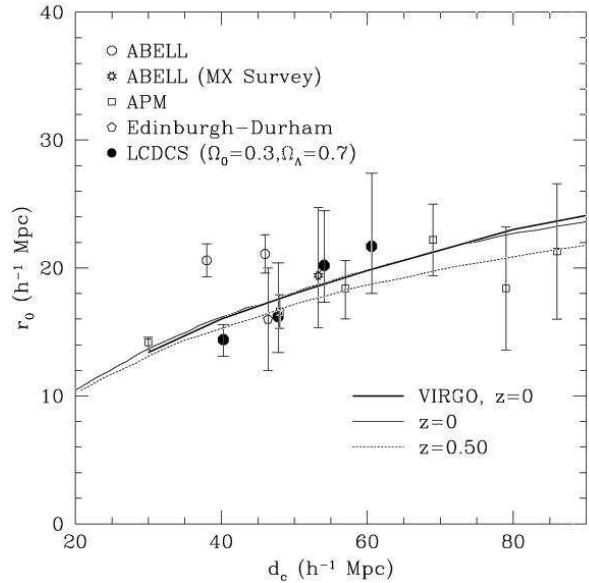}{200 pt}{0}{40}{40}{-120}{-20}
\caption{Correlation scale-length vs. mean cluster separation. 
Solid line represents results from VIRGO simulations at $z = 0$
and the dotted line represents theoretical expectation for $z = 0.5$.}
\end{figure}

The Las Campanas Distant Cluster Survey provides another view on
cluster selection and so complements not only other optical surveys,
but surveys at other wavelengths. Every method has potential
systematic problems, and rather than promoting one method over 
the others, we encourage cross-comparison of the various methods
to identify and resolve those problems. To that aim, 1) we have published
our catalog (Gonzalez {\it et al.} 2001), 2) we have targeted 18
X-ray clusters and recovered all but one, which is at low Galactic latitude,
thereby demonstrating that our false negative rate with respect
to bona-fide X-ray clusters is low, 3) begun a weak-lensing analysis
of 20 of our candidates (the only one analyzed to date by D. Clowe shows 
a significant mass signature, as does a less massive candidate cluster
in the same field) and 4) are in the process of obtaining Chandra data
for four candidates. SZ observations of some of these clusters would
add significantly to our understanding of the selection function.

The study of clusters has entered a new era where large samples
of candidate clusters are becoming common. It becomes increasingly important
to acknowledge that the detailed studies of clusters as done
when only a few clusters where available 
does not fully exploit the power of these
large samples. We have begun to develop redshift and mass estimators
from our survey data and apply them to produce statistical results
obtained from the entire catalog. Future work must focus on refining
these calibrations. We are no longer limited by cluster statistics, 
but we remain limited by systematic uncertainties.

\acknowledgments

DZ acknowledges financial support from
the David and Lucile Packard Foundation, 
the Sloan Foundation, the NSF CAREER program (AST 97-33111),
and the conference organizers. Data fron the ESO distant cluster survey
obtained from the ESO NTT and VLT telescopes.


\end{document}